\documentclass[aps,11pt, superscriptaddress, showpacs,showkeys,floatfix]{revtex4}

\usepackage[english]{babel}

\usepackage{graphicx}
\usepackage{colordvi}

\usepackage{amsmath}
\usepackage{amsfonts}
\usepackage{amssymb}
\usepackage{stfloats}
\usepackage{color}
\usepackage{graphicx}

\begin{document}

\title{Primordial Universe with radiation and Bose-Einstein condensate}

\author{R. C. Freitas\footnote{E-mail: rodolfo.freitas2@fatec.sp.gov.br}}

\affiliation{Faculdade de Tecnologia de Jundia\'{\i} "`Deputado Ary Fossen"',\\ 
Av. Uni\~ao dos Ferrovi\'arios, CEP 13201-160, Jundia\'{\i}, SP, Brazil.}

\author{G. A. Monerat\footnote{E-mail: monerat@uerj.br}}

\affiliation{Departamento de Modelagem Computacional, 
Instituto Polit\'{e}cnico, \\
Universidade do Estado do Rio de Janeiro, CEP 28.625-570, Nova Friburgo - RJ - Brazil.}
 
\author{G. Oliveira-Neto\footnote{E-mail: gilneto@fat.uerj.br}}

\affiliation{Departamento de F\'{\i}sica, Instituto de Ci\^{e}ncias Exatas, 
Universidade Federal de Juiz de Fora, \\
CEP 36036-330, Juiz de Fora, Minas Gerais, Brazil.}

\author{F. G. Alvarenga\footnote{E-mail: flavio.alvarenga@ufes.br}}

\author{S. V. B. Gon\c{c}alves\footnote{E-mail: sergio.vitorino@pq.cnpq.br}}

\author{R. Fracalossi \footnote{E-mail: rfracalossi@gmail.com}}

\affiliation{Departamento de F\'{\i}sica, Centro de Ci\^{e}ncias Exatas,\\
Universidade Federal do Esp\'{\i}rito Santo, CEP 29075-910, Vit\'{o}ria, ES, Brazil.}

\author{E. V. Corr\^{e}a Silva\footnote{E-mail: evasquez@uerj.br}}
 
\author{L. G. Ferreira Filho\footnote{E-mail: kph120@gmail.com}}

\affiliation{Departamento de Matem\'{a}tica e Computa\c{c}\~{a}o, 
Faculdade de Tecnologia, \\ 
Universidade do Estado do Rio de Janeiro, CEP 27523-000, Resende-RJ, Brazil.}

\date{\today}

\begin{abstract} \noindent In this work we derive a scenario {in which} the early universe consists of {radiation fluid} and Bose-Einstein condensate. The possibility of gravitational self-interaction due to an attractive Bose-Einstein condensate is analyzed. The classical behavior of the scale factor of the universe is determined by a parameter associated with the Bose-Einstein fluid with bouncing or Big Crunch solutions. After we proceed to compute the finite-norm wave packet solutions to the Wheeler-DeWitt equation. The behavior of the scale factor is studied by applying the many-worlds interpretation of quantum mechanics. The quantum cosmological model is free from the singularities.
\end{abstract}



\maketitle

\section{Introduction}

The study of the primordial evolution of the universe is one of the greatest challenges of modern science. Also known as Planck's era, this initial period consist, in addition with a space-time singularity, an inflationary phase followed by a reheating to a situation where the emerging universe would have a size of approximately $10^{-5}$ m. All this evolution happening in a tiny fraction of a second. This description is made by joining the cosmological principle of isotropy and homogeneity, the Hubble law, and the field equations of General Relativity in the so-called Big Bang Theory or Standard Cosmological Model (SCM). But, still according to the SCM, the existence of this initial singularity produces fundamental inconsistencies in the description of the first moments in which the universe emerges. This situation is considered as a failure in the initial description of the universe and it seems to be a consensus that we need a quantum theory of gravitation to understand these early moments. This theory with observational results to prove it does not yet exist in a consistent way and many questions are still unanswered about this initial phase. Quantum cosmology \cite{Halliwell, Moniz} can provide elements of the early universe, as a good toy model. It is not a simple framework but with some hyphotesis we can obtain informations about this primordial era \cite{bojo, kie, bojo01}. On the other side, the SCM explains observations consistently in a simple framework but has other problems \cite{Kolb, barbara, prana, arm, mico}.

\par
We can consider the Bose-Einstein condensate (BEC) to explain the origin and nature of DM \cite{harko1, harko3, freitas.bec, chavanis1}. The BEC process, that is a well observed phenomenon in terrestrial experiments, occurs when a gas of bosons is cooled  {down to} very low temperatures, near absolute zero, {thus making} a large fraction of the particles occupy the same ground state. The BEC model can also be applied to cosmology in order to describe the evolution of the recent universe \cite{eck}. In this attempts it can be assumed that this kind of condensation could have occurred at some moment during the  {cosmic history.} The cosmic BEC mechanism was broadly discussed
in \cite{bec1, bec2}. In general the BEC takes place when the gas temperature is below the critical temperature
{$T_{crt}=2\pi\hbar^{2}n^{2/3}/mk_{B}$,} where $n$ is the {particle} density, $m$ is the particle mass and $k_B$ is the Boltzmann's constant.
Since in an adiabatic process the matter dominated universe behaves as $\rho\propto T^{3/2}$ the cosmic dynamics has the same temperature
dependence. Hence we will have the critical temperature at present $T_{crt}=0.0027~\textrm{K}$ if the boson temperature was equal to the
radiation temperature at the redshift $z=1000$. During the cosmic adiabatic evolution the ratio of the photon temperature and the matter
temperature evolves as $T_{r}/T_{m}~\propto~a$, where $a$ is the scale factor of the universe. Using as value for the present energy density of
the universe $\rho=9.44\times10^{-30} \textrm{g}/\textrm{cm}^{3}$ BEC will {happen} if the boson mass satisfies $m<~1.87~\textrm{eV}$.
\par
The Gross-Pitaevskii (GP) equation is a long-wavelength theory widely used to describe dilute BEC. Since the GP equation fails to describe short-ranged repulsive interactions in low dimensions \cite{modbec}, the ground state features of the BEC can be described by a generalized GP equation \cite{bec3, harko2}, where the inter-particle interaction term in the GP equation is modified. This generalized GP equation is
\begin{equation}
   \label{eq.gGP}
   \dot{\imath}\hbar\frac{\partial \phi(t,\vec{r})}{\partial t}=-\frac{\hbar^2}{2m}\nabla^2\phi(t,\vec{r})+mV(\vec{r})\phi(t,\vec{r})+g'(n)\phi(t,\vec{r}) \quad ,
\end{equation}
where $\phi(t,\vec{r})$ is the wave function of the condensate, $m$ is the particles mass, $V$ is the gravitational potential that satisfies the Poisson's equation $\nabla^2V(\vec{r})=4\pi G\rho$, $g'=dg/dn$, $n=|\phi(t,\vec{r})|^2$ is the BEC density and $\rho=mn$. We use the Madelung representation of the wave function in order to understand the physical properties of a BEC, which means that
\begin{equation}
   \phi(t,\vec{r})=\sqrt{n(t,\vec{r})}\times e^{\dot{\imath}S(t,\vec{r})/\hbar} \quad ,
\end{equation}
where $S(t,\vec{r})$ has the dimension of an action. This transformation above will make the generalized GP Eq. (\ref{eq.gGP}) to break into two equations
\begin{eqnarray}
   \frac{\partial \rho}{\partial t} +\nabla \cdot(\rho \vec{v}) & = & 0 \quad , \\
   \rho \left(\frac{\partial \vec{v}}{\partial t}+(\vec{v}\cdot \nabla)\vec{v}\right) & = & -\nabla p\left(\frac{\rho}{m}\right)-\rho\nabla\left(\frac{V}{m}\right) -\nabla V_Q \quad ,
\end{eqnarray}
where $V_Q=-(\hbar^2/2m)\nabla^2\sqrt{\rho}/\rho$ and $\vec{v}=\nabla S/m$ is a quantum potential and the velocity of the quantum fluid, respectively. The effective pressure of the condensate is defined as
\begin{equation}
   p\left(\frac{\rho}{m}\right)=g'\rho -g \quad .
\end{equation}
If we write $g \propto \rho^\gamma$, one can find the generalized equation of state (EoS)
\begin{equation}
   \label{eq.eos2}
	  p = \sigma \rho^\gamma \quad,
\end{equation}
where $\sigma$ is a proportionality constant that will be determined in the context of our model and can be related to the mass and the scattering length of the boson in the long-wavelength theory, and $\gamma\equiv 1+1/n$ is the polytropic index.

We can generalize the BEC EoS (\ref{eq.eos2}) even further \cite{chavanis4}
\begin{equation}
p = \omega\rho + \sigma\rho^{1 + 1/n}\quad,
\end{equation}
to describe the physical state of the matter content of the universe, as the sum of a standard linear EoS and the polytropic term.

Here we will describe a specific model of particular physical interest \cite{chavanis2, chavanis3}, where $n$ is chosen equal to unity, so that the polytropic term correspond to ordinary BEC. This EoS is written as
\begin{equation}
p=\omega\rho+\sigma\rho^2\quad,
\label{eq1}
\end{equation}
where the polytropic constant $\sigma$ represents a self-interaction and the $\omega$ represents the linear term with $-1\leq \omega \leq 1$, where $\omega=1/3$ is radiation, $\omega=0$ is dust matter, $\omega=-1$ is cosmological constant and the less known stiff matter is described by $\omega=1$ \cite{zeldo}. 

At late times, when the density is low, the BEC contribution to the EoS is negligible and the evolution is determined by the linear term. But in the early universe, when the density is high and $(1+\omega+\sigma\rho^{2}) > 0$, the term due to BEC in the EoS is dominant and modifies the dynamics of the universe. Lately this model was used as a model of the early universe. We can assume that this generalized EoS holds before radiation era and for the case of attractive self-interaction the universe has always existed and for the non-physical limit $t\rightarrow-\infty$ the density tends to a constant value and the radius goes to zero, both exponentially \cite{chavanis2, chavanis3}. 

For the  EoS equation (\ref{eq1}) with $\omega \neq -1$ the energy conservation equation is
\begin{equation}
   \label{eq:conservacao}
   \dot{\rho}+3H\rho(1+\omega+\sigma\rho)=0 \quad,
\end{equation}
where dot denotes a derivative with respect to the cosmic time $t$ and $H=\dot{a}/a$ is the Hubble parameter. This equation can be easily integrated to give
\begin{equation}
\label{eos01}
   \rho = \frac{\rho_{*}}{(a/a_0)^{3(1+\omega)} + 1} \quad,
\end{equation}
where $a_0$ is a constant of integration, and $\rho_{*}=\left(1+\omega\right)/|\sigma|$.

In the case of an attractive self-interaction ($\sigma<0$) the density is defined for $0<a<\infty$, and
\begin{equation}
\left\{
\begin{array}{lll}
   \frac{\rho}{\rho_{*}}\approx
	      \quad \quad \quad 1 \quad \quad \quad , \quad a \rightarrow  0 \quad, \\
				(a/a_0)^{-3(1+\omega)} \rightarrow 0 \quad , \quad a \rightarrow \infty
		  \quad ,
\end{array}
\right.
\end{equation}
with, in the same limits, $p=-\rho_{*}$ and $p\rightarrow 0$.

In the present paper we will study the dynamics of a primordial universe filled with BEC and a radiation perfect fluid {{($p_{rad}=\alpha\rho_{rad}$ with $\alpha=1/3)$}}, using quantum cosmology. The universe has a Friedman-Lemaitre-Robertson-Walker (FLRW) geometry and the spatial sections have constant positive curvatures. In particular, we want to determine if the quantum description removes the singularities present
in the classical model. The classical model with an attractive self-interaction exhibits a class of singular solutions for certain values of the polytropic constant $\sigma$ and initial conditions. At quantum level, the Galerkin's spectral method \cite{galerkin} is used for approximate calculation of eigenvalues and eigenfunctions of the Wheeler-DeWitt equation (an unidimensional Schr\"odinger-like equation, for this model). Wave packets will be constructed and expected values will be calculated employing the interpretation of many-worlds of quantum mechanics. The results reveal the non-singular quantum universe.

The paper is organized as follows.  In Sec. (\ref{CM}) we apply the Schutz's formalism \cite{Schutz} in a classical FLRW cosmological model with perfect fluid and condensate Bose-Einstein. The evolution of the scale factor of the universe is analysed.  We present next, Section (\ref{CQ}), the quantum model that we will work obtaining the equation that drives the dynamics of the scenario, the Wheeler-DeWitt equation. Wave-packet solutions to the Wheeler-DeWitt are found for the gravitational attractive self-interaction and expectation values for the scalar factors are evaluated.  Sec. (\ref{CC}) constitutes  a summary of the results herein presented.

\section{The Classical Model}
\label{CM}

We use the Friedmann-Lemaitre-Robertson-Walker metric to describe the primordial phase of the universe, written as
\begin{equation}
ds^2=-N^2(t)dt^2+a^2(t)\left(\frac{dr^2}{1-k r^2}+r^2d\Omega^2\right)\quad,
\label{sec2eq1}
\end{equation}
where $a(t)$ is the scale factor, $d\Omega$ is the line element of the 2D unit sphere, $N(t)$ is the lapse function and $k$ determines the curvature of the spatial geometry of the universe and indicates whether the universe is open ($k = -1$), closed ($k - +1$) or flat ($k = 0$). Here we use natural units where $\hbar = 1$ and $8\pi G = c =1$. In our model the matter content of the universe will consist of the perfect fluid with a barotropic equation of state (EoS) $p = \omega\rho$ ($-1 < \omega < +1$) plus the condensate Bose-Einstein, described by the EoS (\ref{eq1}). The energy momentum tensor is defined as
\begin{equation}
T_{\mu\nu}= (\rho+p)U_{\mu}U_{\nu} - pg_{\mu\nu}\quad,
\label{sec2eq2}
\end{equation}
where $U_{\mu}$ is the four-velocity, $\rho$ the energy density, $p$ the pressure of the fluid and $g_{\mu\nu}$ represents the metric tensor. This tensor characterizes an isotropic fluid in comoving coordinates providing the 4-velocity as timelike where $U^{\mu} = \delta^{\mu}_{0}$.
{Moreover, a radiative fluid with  equation state of the form $p_{rad}=\frac{1}{3}\rho_{rad}$  has been included in the model; its
energy-momentum tensor has the same form as (\ref{sec2eq2}). Only the case of positive curvature $(k=1)$ of the spatial section is
considered in this work.}

The classical dynamical of the system is given by the metric (\ref{sec2eq1}) and the energy momentum tensor (\ref{sec2eq2}). We assume that those two fluids do not interact with each other. Such a procedure is essentially equivalent to consider the Hamilton's equations derived from a total Hamiltonian ${\cal H}=N(t)H$, such that \cite{Julio}
\begin{equation}
{\cal H}=-\frac{p_{a}^{2}}{24}-6a^2+a^4 p_{poly}+p_{T}\quad.
\label{sec2eq3}
\end{equation}

\noindent
Here, $p_{a}$ and $p_{T}$ are the canonical momenta conjugated to the scale factor $a(t)$ and the variable $T$,
that describes the radiation fluid, respectively. We also have $N(t)=a(t)$. The polytropic pressure $p_{poly}$ is given by
\begin{equation}
{p_{poly}}=- \frac{|\sigma|^2}{\left(a^{3}+ 1\right)^2}\quad.
\label{sec2eq4}
\end{equation}
In order to obtain Eq. (\ref{sec2eq3}), the authors of Ref. \cite{Julio}
used the Schutz's variational formalism which describes a relativistic fluid interacting with the gravitational field and endows the fluid with dynamical degrees of freedom \cite{Schutz, Furtado}.

Using the Hamilton's equations, we can write
\begin{equation}
\left\{
\begin{array}{lllllll}
a' =& \frac{\partial {\cal H}}{\partial p_{a}}=-\frac{p_{a}}{12}\quad,\\
& \\
p_{a}' =& - \frac{\partial {\cal H}}{\partial a}=12a-4a^3p_{poly}(a)-\frac{6|\sigma|^2a^6}{(a^3+1)^3}\quad, \\
&\\
T' =& \frac{\partial {\cal H}}{\partial p_{T}}=1\quad, \\
& \\
p_{T}' =& - \frac{\partial {\cal H}}{\partial T}=0\quad, \\
\end{array}
\right.
\label{sec2eq5}
\end{equation}
where prime indicates a derivative with respect to the conformal time $\eta$, with $d\eta \equiv a(t)dt$.

By analyzing the set of pairs $(a, p_{a})$ of the phase space we find that all solutions are limited. The universe expands initially reaching a maximum value and then the contraction begins. We have two different kinds of situations, according to the value of $\sigma$: (i) bouncing and Big-Crunch solutions (that depends on the initial conditions) or (ii) only Big-Crunch solutions. 
\begin{figure}[h!]
\begin{center}
\begin{minipage}{0.45\linewidth}
\includegraphics[height=5.0cm,width=5.0cm]{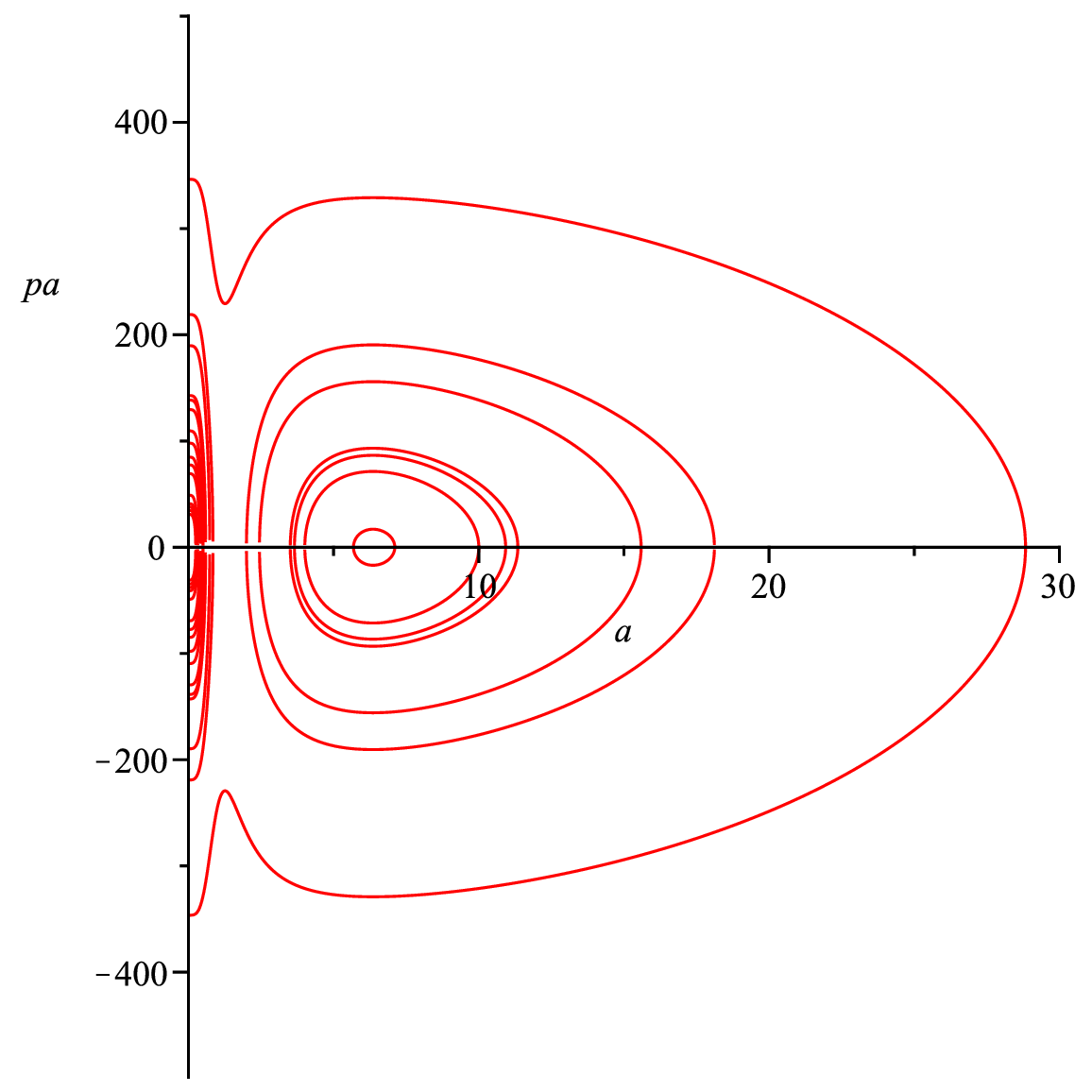}
\centerline{(a)}
\end{minipage}
\begin{minipage}{0.45\linewidth}
\includegraphics[height=5.0cm,width=5.0cm]{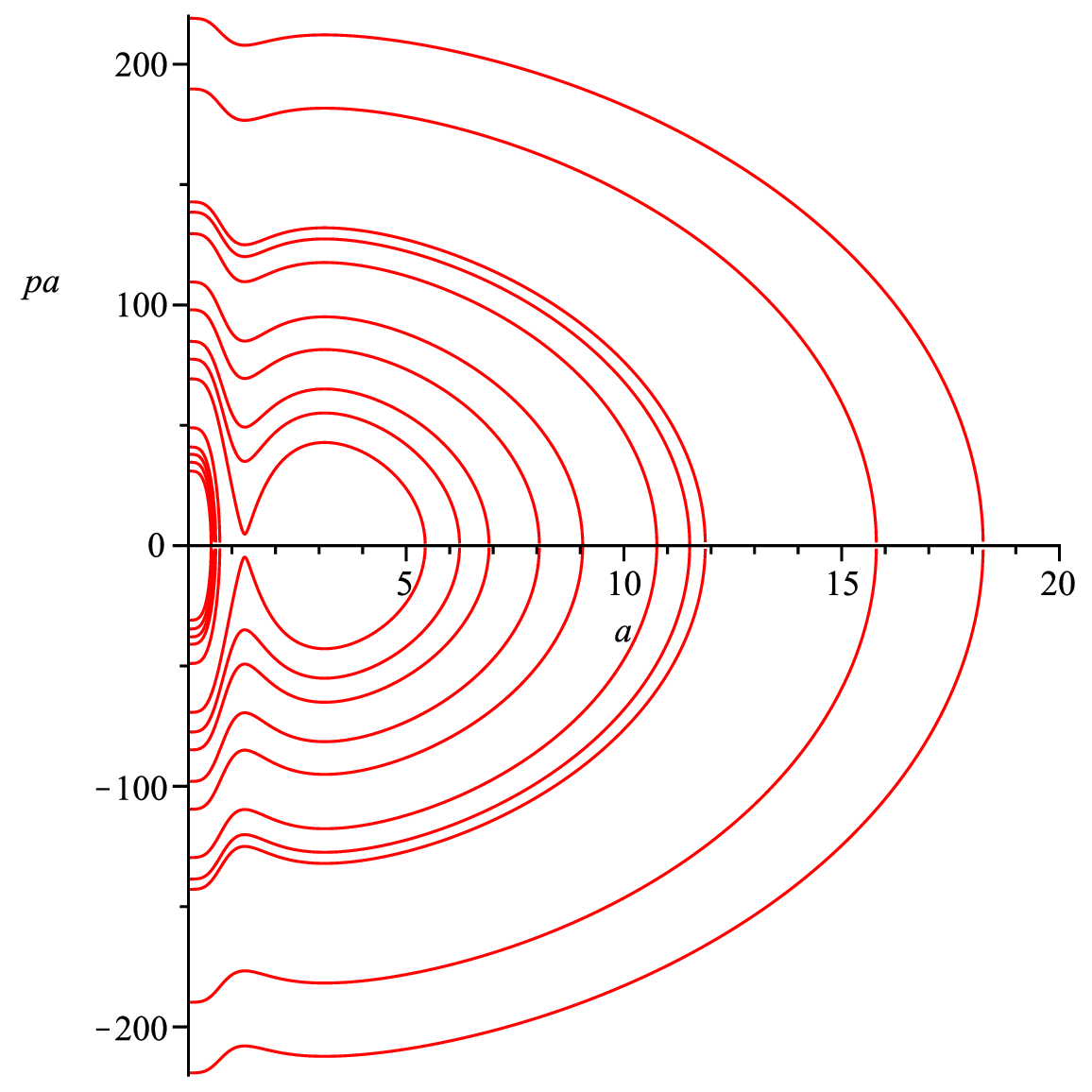}
\centerline{(b)}
\end{minipage}
\end{center}
\caption{{{The curves of pairs $(a, p_{a})$ represent homogeneous and isotropic universes with positive curvature ($ k = 1 $), a radiation fluid and Bose-Einstein condensate with attractive gravitational self-interaction ($\sigma < 0$). (a) For $\sigma = -100$ we observe that in addition to cases in which the universe evolves into a Big-Crunch, there are a bouncing solutions for some initial conditions. (b) For $\sigma = - 26$ we see that there are no bouncing solutions.}}}
\label{fricochete}
\end{figure}

The presence of one or another solution depends on the values assumed by the parameter $\sigma$ related to the Bose-Einstein condensate. The study showed that if we consider only integer values for the parameter $\sigma$ we will have bouncing solutions in the range $-1020 \leq\sigma\leq -27$. For values outside this range such solutions disappear, remaining only solutions with Big Crunch. The Figures \ref{fricochete} (a) and \ref{fricochete} (b) show each of these cases, respectively.
		
When we look at the Figure \ref{fricochete} (a) we see that, depending on the initial conditions, we have the possibility of finding not only solutions whose final configuration is collapsing towards singularity in $a = 0$, but also the possibility to find bouncing solutions. On the other hand, in the Figure \ref{fricochete} (b), we observe that whatever initial conditions are chosen, after a time interval the universe will collapse towards the initial singularity at $a = 0$, resulting in a Big-Crunch.

The classical evolution of the scale factor of the universe (in conformal time) is obtained by combining the equations (\ref{sec2eq5}) 

\begin{eqnarray}
{\frac {d^{2}}{d{\eta}^{2}}}a \left( \eta \right) +a \left( \eta \right) +\frac{|\sigma|^2}{3}\,{
\frac { \left( a \left( \eta \right)  \right) ^{3}}{ \left(  \left( a
 \left( \eta \right)  \right) ^{3}+1 \right) ^{2}}}-\frac{|\sigma|^2}{2}\,{\frac { \left( a
 \left( \eta \right)  \right) ^{6}}{ \left(  \left( a \left( \eta \right) 
 \right) ^{3}+1 \right) ^{3}}}=0\quad,
\label{eqFriedmann}
\end{eqnarray}

Solving numerically the Eq. (\ref{eqFriedmann}) for the values of $\sigma$ shown in the Figures \ref{fricochete} (a) and (b), we can confirm the cosmological predictions described in the previous paragraph. 

\begin{figure}[h!]
\begin{center}
\begin{minipage}{0.45\linewidth}
\includegraphics[height=5.0cm,width=5.0cm]{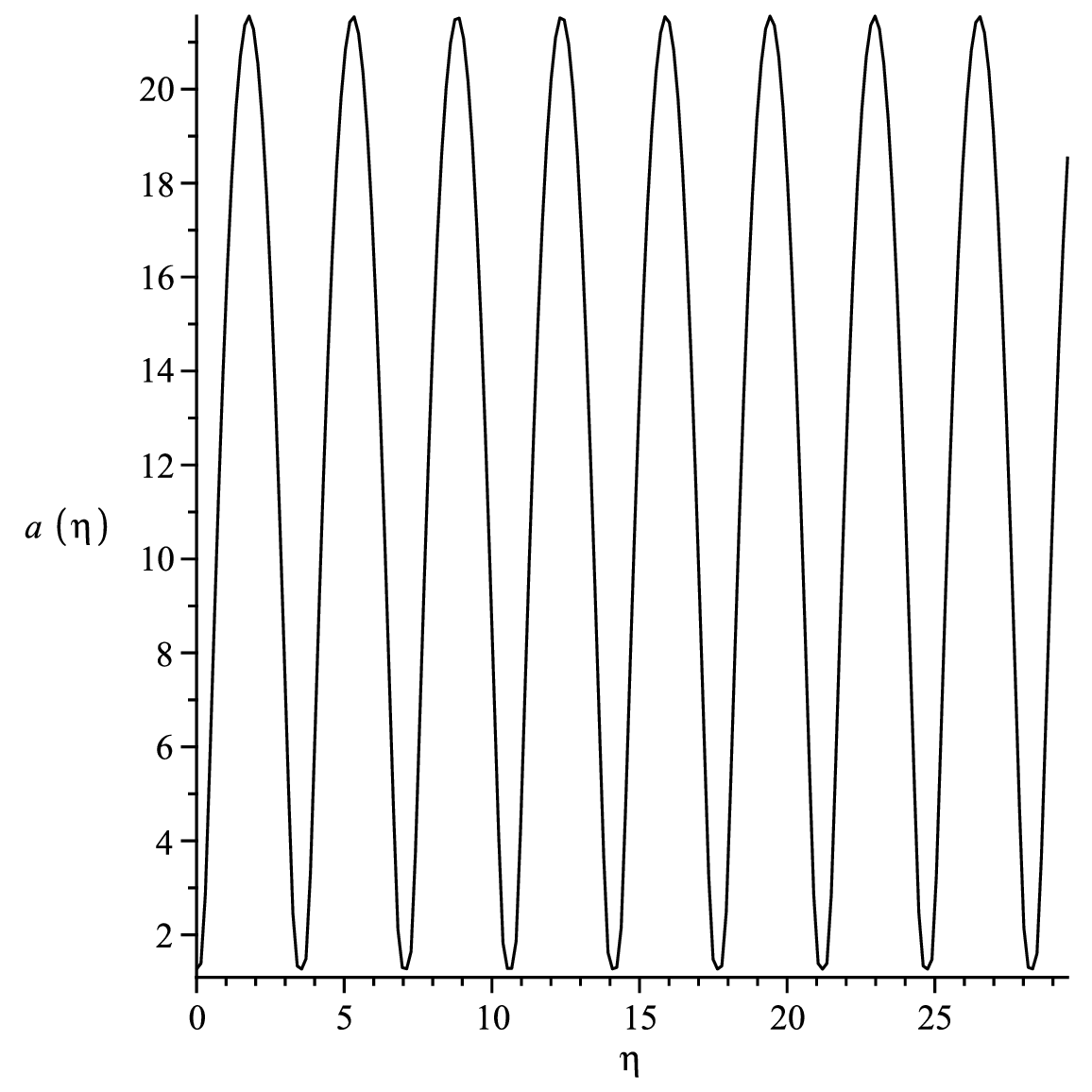}
\centerline{(a)}
\end{minipage}
\begin{minipage}{0.45\linewidth}
\includegraphics[height=5.0cm,width=5.0cm]{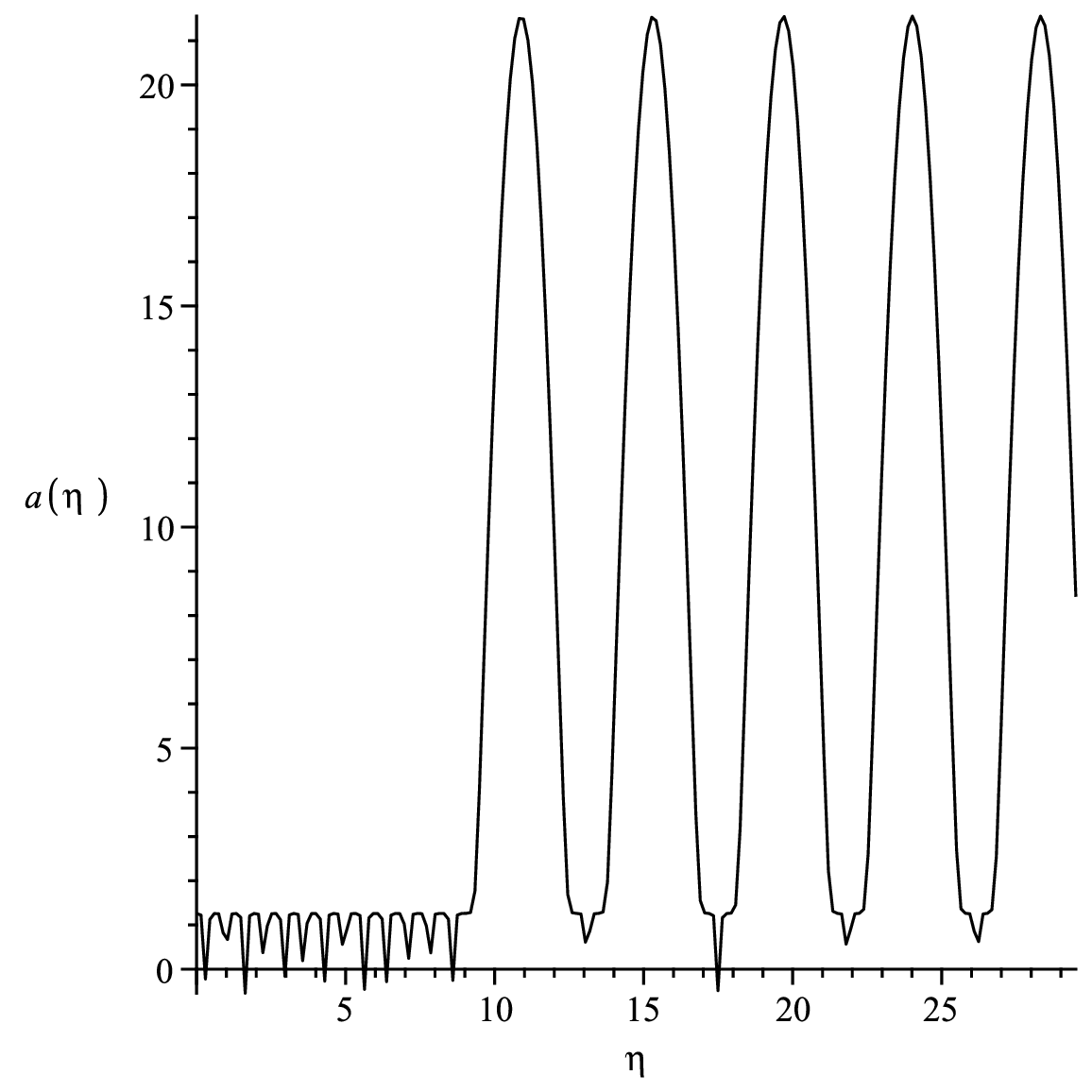}
\centerline{(b)}
\end{minipage}
\end{center}
\caption{{{Behaviour of the scale factor of the FLRW universe with $k = 1$ in the case of radiation fluid and Bose-Einstein condensate with attractive gravitational self-interaction ($\sigma = -100$). In (a), for the initial conditions $a(0) = 1.27$ and $a^{\prime}(0) = 0$, we have an oscillating  universe between a minimum and a maximum without the presence of singularities. In (b), for initial conditions $a(0) = 1.26$ and $a^{\prime}(0) = 0$, after a short time interval the universe collapses toward the initial singularity $a = 0$, leading to a Big Crunch.}}}
\label{fig0}
\end{figure}

The Figure \ref{fig0} shows the behavior of the time evolution of the scale factor for the Bose-Einstein condensate parameter of $\sigma = -100$. The initial conditions considered in Figure 2 (a) are $a(0) = 1.27$ and $a^{\prime}(0)=0$. The result shows a universe free of singularity, with bouncing solutions. If we adopt initial conditions $a(0) = 1.26$ and $a^\prime(0) = 0$ we reveal a universe that in a short time interval collapse towards the singularity $a = 0$. Such behaviors are in accordance with Figure 1 (a). For this specific value of $\sigma=-100$, if we consider the initial conditions $a(0)=a_{0}$ and $a^{\prime}(0)=0$, we will have solutions without singularities for $1.27 \leq a_{0} \leq 21.55$. 

\begin{figure}[h!]
\begin{center}
\begin{minipage}{0.45\linewidth}
\includegraphics[height=5.0cm,width=5.0cm]{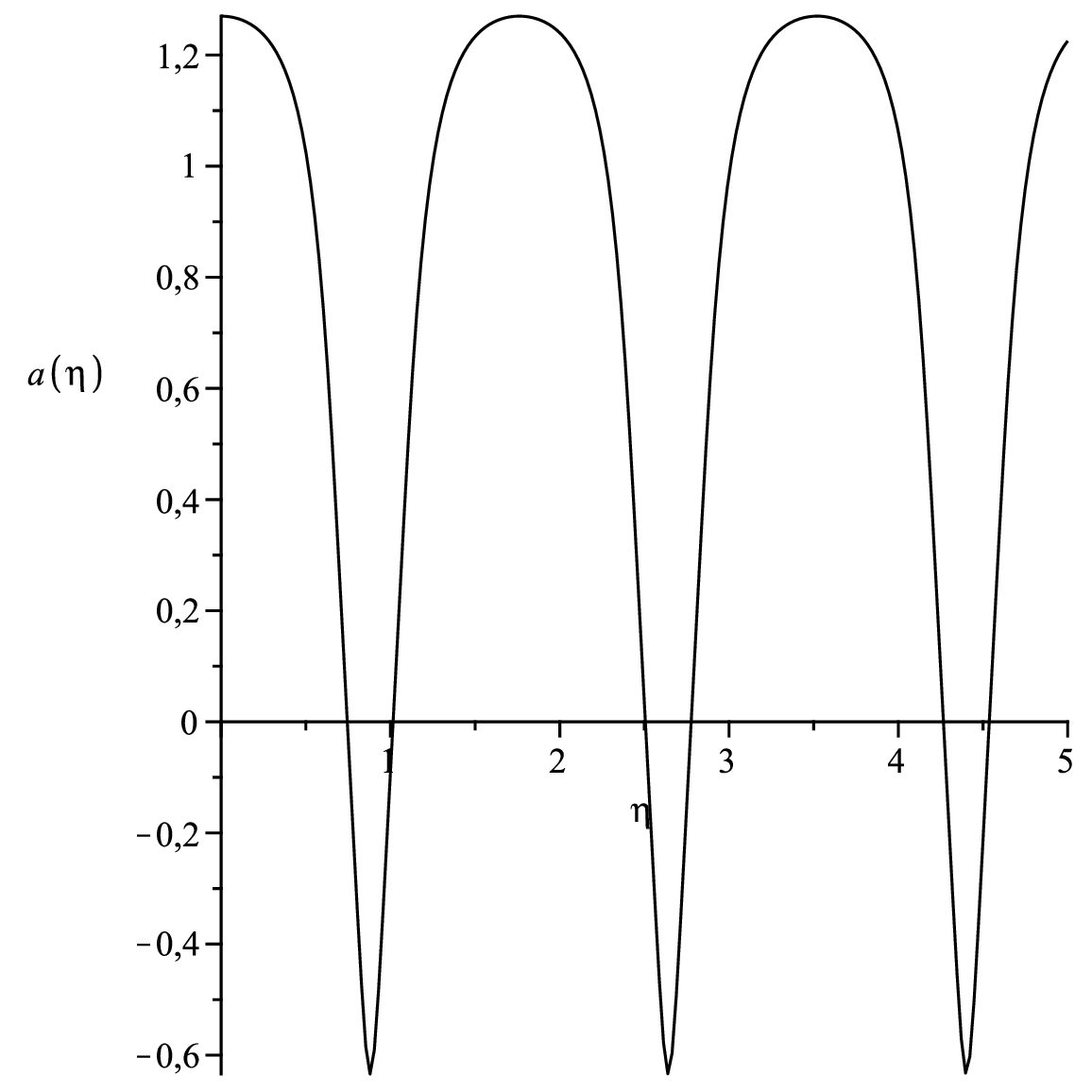}
\centerline{(a)}
\end{minipage}
\begin{minipage}{0.45\linewidth}
\includegraphics[height=5.0cm,width=5.0cm]{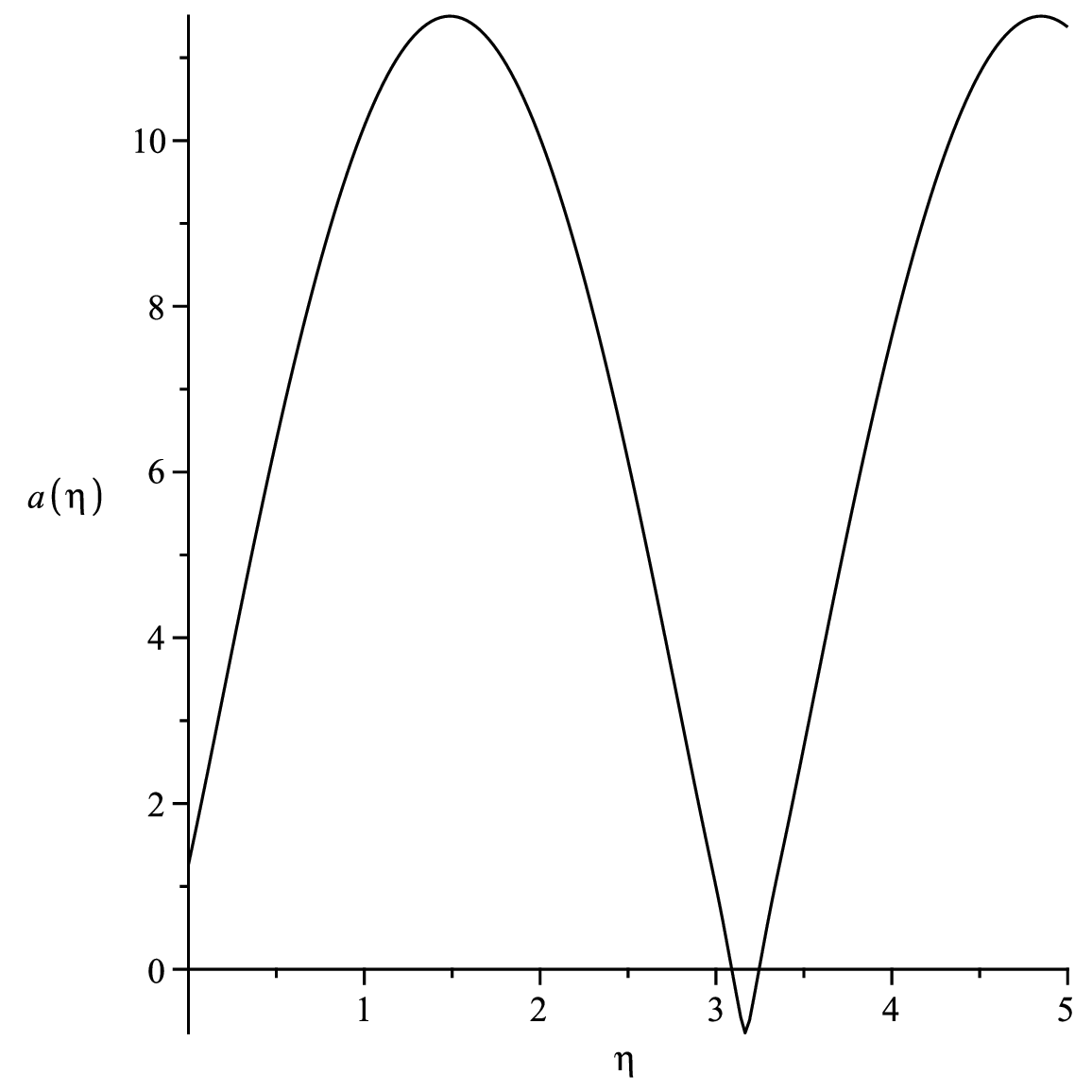}
\centerline{(b)}
\end{minipage}
\end{center}
\caption{Behavior of the scale factor of the FLRW universe with $k = 1$ in the case of a radiation fluid and a Bose-Einstein condensate with attractive gravitational self-interaction ($\sigma = -26$). Here both results show after a short time interval a collapsing universe toward the initial singularity at $a = 0$. The initial conditions considered here were: (a) $a(0) = 1.27$ and $ a^{\prime}(0) = 0$ and (b) $a(0) = 1.26$ and $ a^{\prime}(0) = 10 $.}
\label{f00}
\end{figure}

If we now consider $\sigma = -26$, regardless of the initial conditions, the result will be a universe collapsing indefinitely, after a certain time interval, toward singularity at $a = 0$. The Figure \ref{f00} shows the time evolution of the scale factor for two sets of initial conditions. This result reflects what appears in the Figure \ref{fricochete} (b).

\section{Canonical Quantization}
\label{CQ}

The quantization follows the Dirac's formalism for constrained systems. For this we introduce a wave function of the canonical variables $a$ and $T$
\begin{equation}
\Psi=\Psi(a,T)\quad.
\label{sec2eq9}
\end{equation}

We impose the appropriate commutations relations between the operators ($\hat a$, $\hat T$) and the respective conjugate momenta ($\hat p_a$, $\hat p_T$). In the Schr\"{o}dinger picture observables are represented by Hermitian operators which act on the wave function: the ``position'' operator $\hat a$ acting on any wave function is equals $a$ multiplied by the same wave function. Their conjugates momenta are represented by the differential operators as
\begin{equation}
\hat{p}_{a}\rightarrow -i\frac{\partial}{\partial a},\,\,\,\,\, \hat{p}_{T}\rightarrow -i\frac{\partial}{\partial T}\quad.
\label{sec2eq10}
\end{equation}

Applying the operator version of the super-Hamiltonian corresponding to equation (\ref{sec2eq3}) we have $\hat{H}\Psi(a, \tau) = 0$. So we obtain the Wheeler-DeWitt equation for this model
\begin{equation}
\left(-\frac{\partial^2}{\partial a^2}+V_{eff}(a)\right)\Psi(a,\tau)=24i\frac{\partial}{\partial \tau}\Psi(a,\tau)\quad,
\label{sec2eq11}
\end{equation}
where we introduce the new variable $T=-\tau$. Here, $V_{eff}(a)$ is the effective potential, which takes the form

\begin{equation}
V_{eff}(a)=144a^2+\frac{24|\sigma|^2a^4}{(a^3+1)^2}\quad.
\label{atrativo-1}
\end{equation}

In the present case, the Eq. (\ref{sec2eq11}) corresponds a time dependent Schr\"odinger-like equation in which the only remaining matter degree of freedom plays the role of time. It is important to say that the $\hat{H}$ must be a self-adjoint operator with respect to inner product \cite{lemos}
\begin{equation}
(\Psi,\, \Psi)=\int_{0}^{\infty} da\; \Psi^{\star}(a,\tau)\Psi(a,\tau)\quad,
\label{sec2eq12}
\end{equation}

\noindent
where the whole structure of Hilbert space is restrict for the set of the wave functions, both 
\begin{equation}
\Psi(0,\tau) =\Psi(\infty,\tau)=0
\label{contorno1}
\end{equation}

\noindent
or 
\begin{equation}
\partial_a\Psi(0,\tau) =\partial_a\Psi(\infty,\tau)=0,
\label{contorno2}
\end{equation}

\noindent
We can solve the Wheeler-DeWitt Eq. (\ref{sec2eq11}) writing $\Psi(a,\tau)$ as
\begin{equation}
\Psi(a,\tau)=e^{-iE\tau}\varphi(a)\quad,
\label{sec2eq13}
\end{equation}

\noindent
where $\varphi(a)$ satisfies the equation
\begin{equation}
\left(-\frac{d^2}{da^2}+V_{eff}(a)\right)\varphi(a)=24E\varphi(a)\quad,
\label{sec2eq14}
\end{equation}

\noindent
with the effective potential described by (\ref{atrativo-1}) and the values of parameters previously defined. We analyze here the gravitational attractive self-interaction case. We will solve Eq. (\ref{sec2eq14}) by using the spectral Galerkin method \cite{galerkin} that proved to be very effective for the quantization of Friedmann-Lemaitre-Robertson-Walker cosmological models with perfect fluid \cite{pedran1, pedran2, monerat21}. For this we will use the SPECTRAL package \cite{monerat22} for the quantization of physical systems.

We call here $\varphi_{p}(a)$ as being the relative eigenmodes to the $p-th$ eigenvalue $E_{p}$ of the Eq. (\ref{sec2eq14}). The Galerkin spectral method states that a possible choice for $\varphi_{p}(a)$ can be written as

\begin{equation}
\varphi_{p}(a)\cong \sum_{n=1}^{\cal{N}}A_{n}^{p}\sqrt{\frac{2}{L}}\sin\left(\frac{n\pi a}{L}\right)
\label{solucoes-autoestados}
\end{equation}

\noindent
{with the coefficients $A_{n}^{p}$ determined by a finite number $M$ of basis functions.

\subsection{Energy Spectrum and eigenfunctions}
\label{espectro}

The energy spectrum was obtained for the first 100 energy levels of the system by the Galerkin spectral method for different values of the Bose-Einstein condensation parameter  $\sigma$. The result of our analysis shows that the lower the value of the parameter $\sigma$, the higher the model energies. We show the first 20 energy levels for five different values of $\sigma$ in the Table \ref{tabela-espectro}.

\begin{table}[h!]{\scriptsize\begin{tabular}{|c|c|c|c|c|c|}
\hline
    &  $\sigma=-1$       & $\sigma=-6$       & $\sigma=-25$        & $\sigma=-50$          & $\sigma=-200$\\ \hline
$m$ &  $E_{m}$           & $E_{m}$           & $E_{m}$             & $E_{m}$               & $E_{m}$\\ \hline
1   & 1.534648581521444  & 1.974485898611541 & 4.067235303918014   & 6.290642383896439     &15.64375886979961\\ \hline
2   & 3.877979846371805  & 5.225359368473296 & 12.02738147723655   & 18.97493977877267     &47.79358127403138\\ \hline
3   & 7.134208459138295  & 8.933629084465441 & 21.44281141036097   & 34.24456145959712     &86.96686525716694\\ \hline
4   & 11.62744603049354  & 13.47279232505126 & 31.76774068919762   & 51.27490671057463     &131.1583600598053\\ \hline
5   & 17.40507584456878  & 19.21359596693284 & 42.70538241125813   & 69.63475284966779     &179.3444791947256\\ \hline
6   & 24.46962015523518  & 26.24713592115603 & 54.05651098337250   & 89.04706735822266     &230.8843067647003\\ \hline
7   & 32.8204972793129   & 34.57744295573775 & 65.67160510140837   & 109.3130682678174     &285.3298861256555\\ \hline
8   & 42.4572262690269   & 44.20042764414275 & 77.4360817974267    & 130.2799947125709     &342.3462521399959\\ \hline
9   & 53.37952381512626  & 55.1132001234115  & 89.30130102567571   & 151.8247517277695     &401.6711222568476\\ \hline
10  & 65.58722520007161  & 67.31405763103633 & 101.4155560426162   & 173.8446005446266     &463.0921144724311\\ \hline
11  & 79.0802315877320   & 80.8019902754355  & 114.2234725590075   & 196.2514098096118     &526.4327928155435\\ \hline
12  & 93.85848137643114  & 95.57637768225672 & 128.2062755110454   & 218.9678796960706     &591.5435914420711\\ \hline
13  & 109.9219347592082  & 111.6368241539612 & 143.5889475727234   & 241.9249364771476     &658.2956300143159\\ \hline
14  & 127.2705651648194  & 128.9830684262605 & 160.4013081741581   & 265.0598600310691     &726.5763388987342\\ \hline
15  & 145.9043543414874  & 147.6149326400957 & 178.6161934878207   & 288.3149045715016     &796.2862682890625\\ \hline
16  & 165.8232894318012  & 167.5322924245567 & 198.2037198218386   & 311.6364099138067     &867.3367004825978\\ \hline
17  & 187.0273611745018  & 188.7350587398216 & 219.1412197744463   & 334.9754650177634     &939.6478238872851\\ \hline
18  & 209.5165627657362  & 211.2231665040159 & 241.4125911518453   & 358.2964594733561     &1013.147310257753\\ \hline
19  & 233.2908891183812  & 234.9965672641276 & 265.0064065164622   & 381.6158662764547     &1087.769187950677\\ \hline
20  & 258.3503363685834  & 260.0552243513123 & 289.9144272704081   & 405.101162056791      &1163.452936790256\\ \hline
\end{tabular}
}
\caption{{\protect\footnotesize {Energy spectrum for the case of radiation and Bose-Einstein condensate with gravitational attractive self-interaction. Here we consider ${\cal N}=100$ and $L=0.8$.}}}
\label{tabela-espectro}
\end{table}

It was possible to obtain the eigenfunctions of the studied system for different values of $\sigma$. Each of the eigenstates obtained here vanishes at $a=0$ and $a=0.8$, satisfying the boundary conditions established by (\ref{contorno1}). In Figure \ref{fautoestadosatrativo}, we consider the first $10$ eigenstates of energy with ${\cal N}=100$ and $L=0.8$.

\begin{figure}[h!]
\begin{minipage}{0.45\linewidth}
\includegraphics[width=0.7\linewidth, height=5cm]{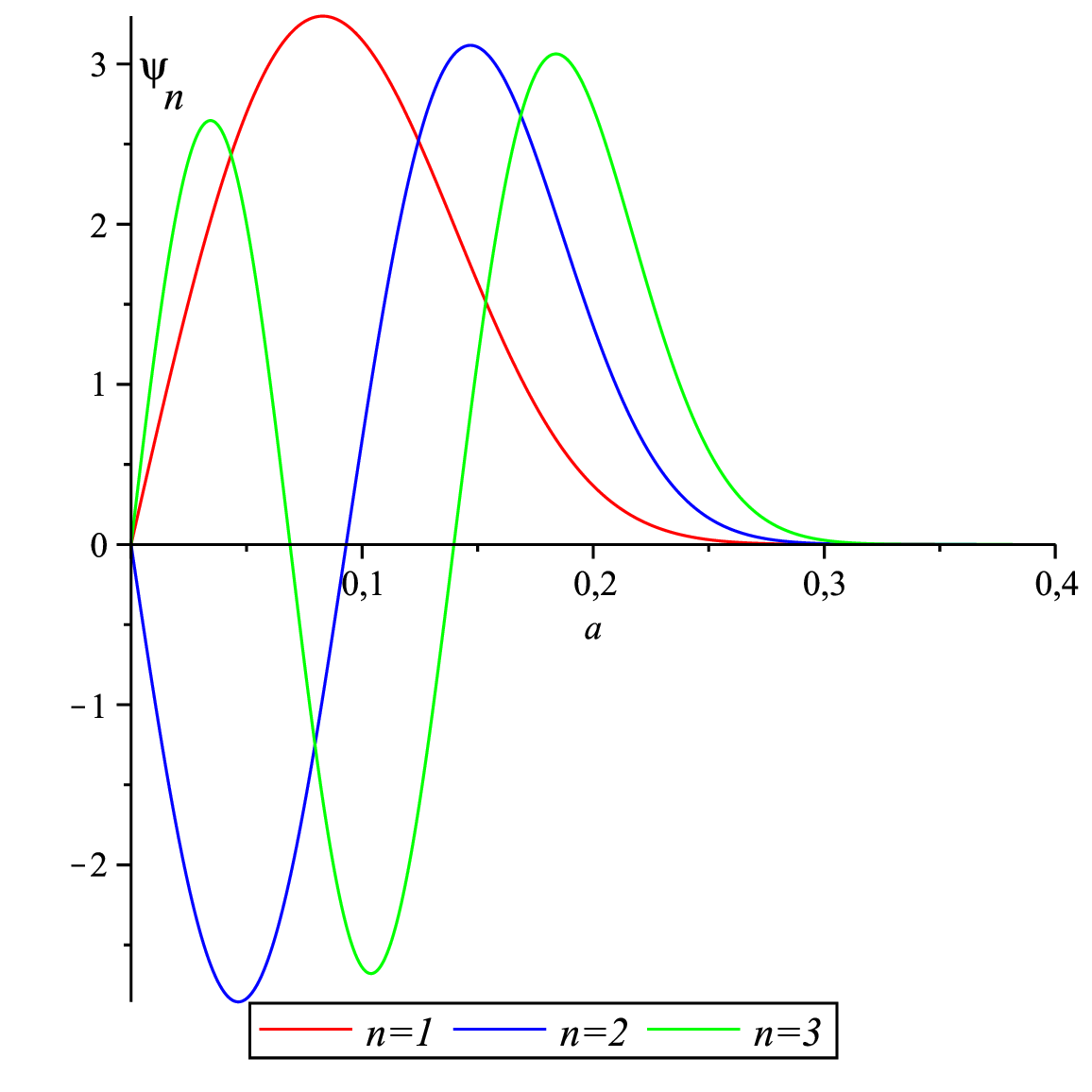} 
\end{minipage}
\begin{minipage}{0.45\linewidth}
\includegraphics[width=0.7\linewidth, height=5cm]{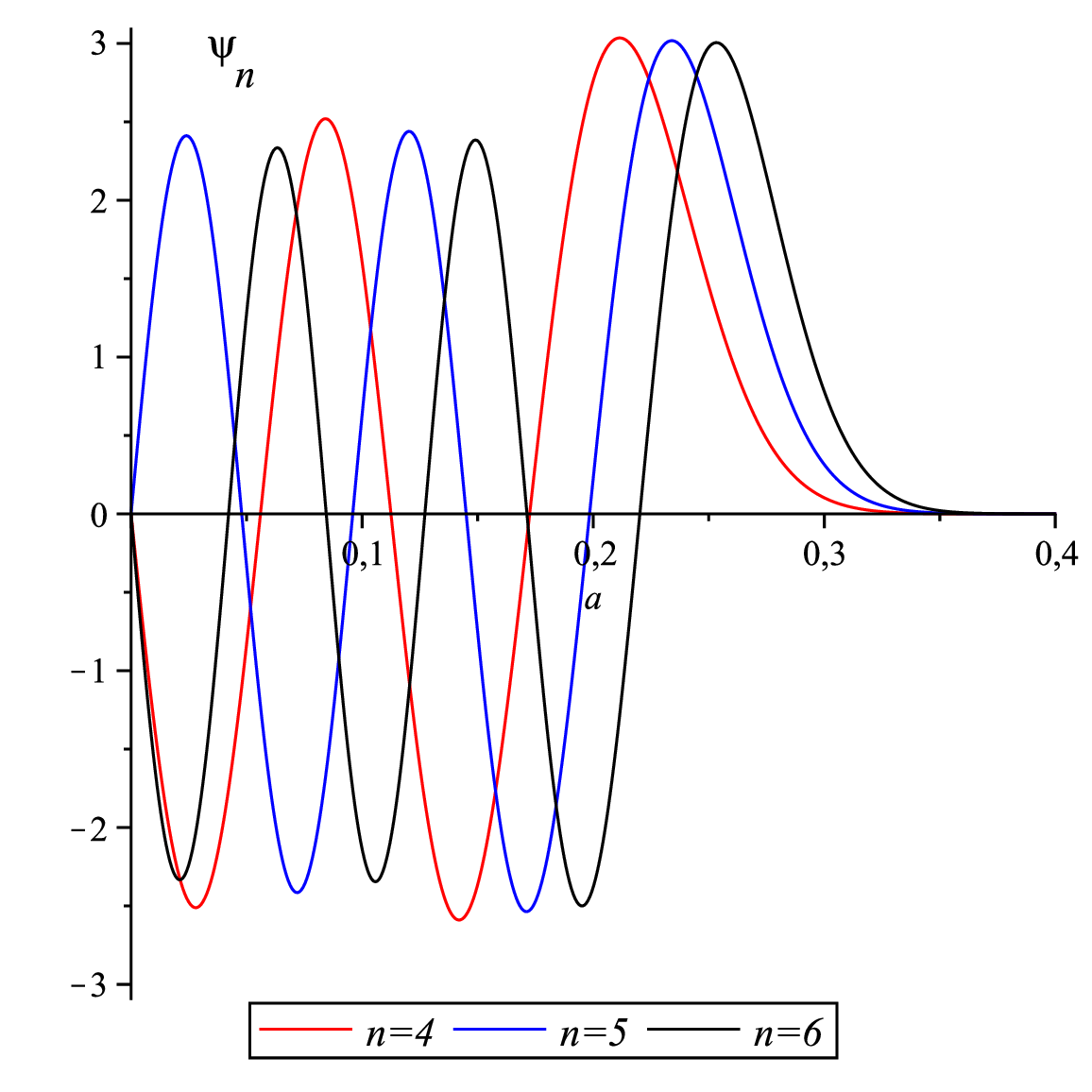}
\end{minipage}
\vspace{0.7cm}
\begin{minipage}{0.45\linewidth}
\includegraphics[width=0.7\linewidth, height=5cm]{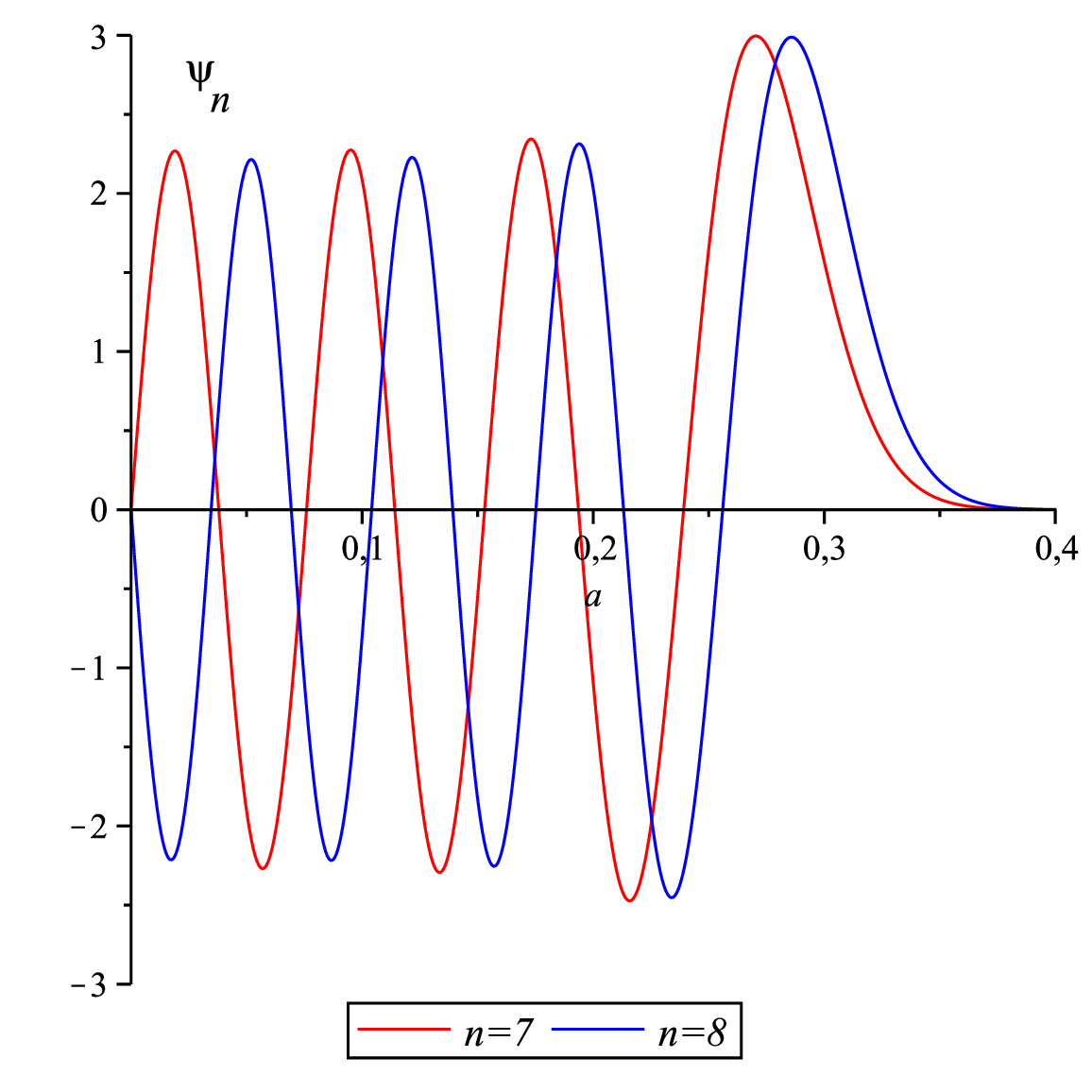} 
\end{minipage}
\begin{minipage}{0.45\linewidth}
\includegraphics[width=0.7\linewidth, height=5cm]{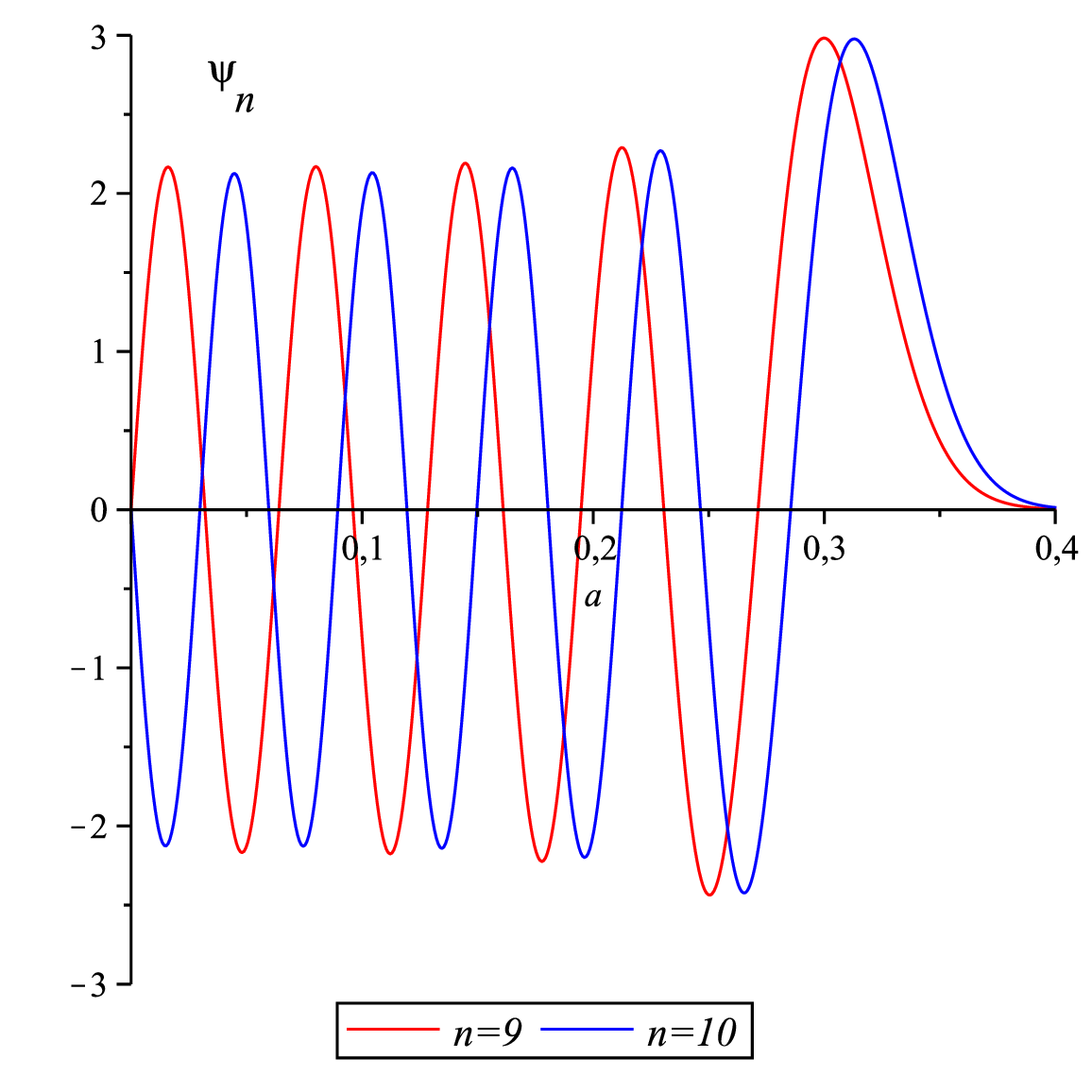} 
\end{minipage}
\vspace{0.1cm}
\caption{The ten lowest energy eigenstates approximate to the case of a Bose-Einstein condensate with gravitational attractive self-interaction and a radiation fluid. Here we consider ${\cal N}=100$, $\sigma=-200$ and $L=0.8$ associated with energy eigenvalues shown in the Table \ref{tabela-espectro}.}
\label{fautoestadosatrativo}
\end{figure}

\subsection{Wave packets, expected values and its uncertainties}
\label{incertezas}

The quantum dynamics of the universe governed by the Wheeler-DeWitt equation occurs through the evolution of wave packets. In our study wave packets are obtained by superposition of a number of their eigenstates. In this work, wave packets will be defined by superposition of the $M$ eigenfunctions chosen from $\cal{N}$ calculated

\begin{equation}
\Psi(a,\tau)=\sum_{p=1}^{ M}\varphi_{p}(a)e^{-iE_{p}\tau}.
\label{pacotes-autoestados}
\end{equation}

 In fact many wave packets were obtained by varying both the value of the Bose-Einstein condensate parameter and the number of superposition eigenstates. In all cases we have obtained a finite norm of the wave packets, well defined throughout the space, even when $a=0$ ($\left[0,L\right]$). It is noteworthy that just like the eigenstates, all wave packets also satisfy the boundary conditions defined in (\ref{contorno1}). The eigenstates have their norm conserved with an accuracy of the order of $10^{-13}$. Tests using the Spectral package also show that the orthogonality among the eigenstates is obtained for the first 20 eigenstates at a precission of $10^{-12}$. In Figure \ref{densidade} we show the probability density distribution at two different time values.

\begin{figure}[h!]
\begin{center}
\begin{minipage}{0.45\linewidth}
\includegraphics[height=5.5cm,width=5.5cm]{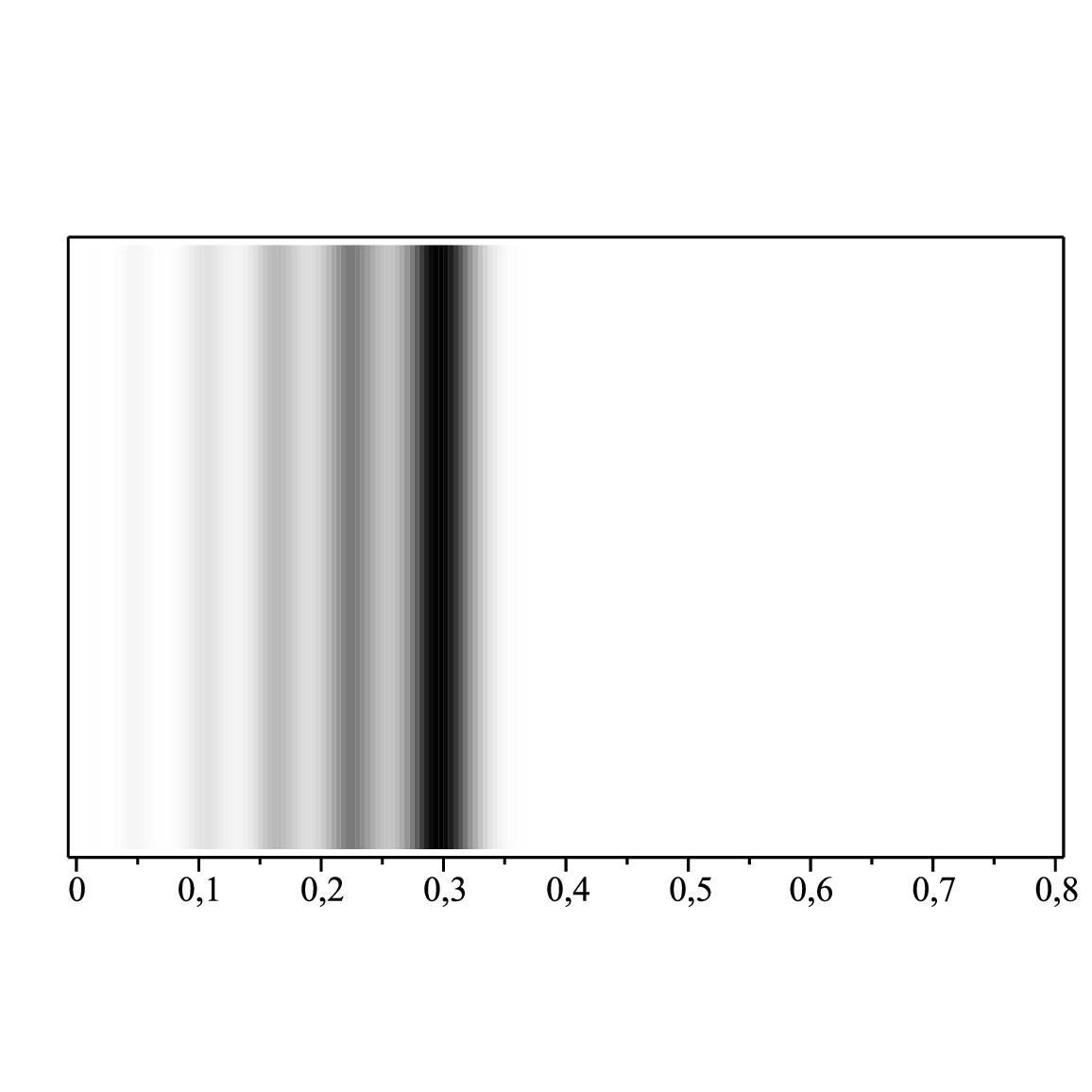}
\centerline{(a)}
\end{minipage}
\begin{minipage}{0.45\linewidth}
\includegraphics[height=5.5cm,width=5.5cm]{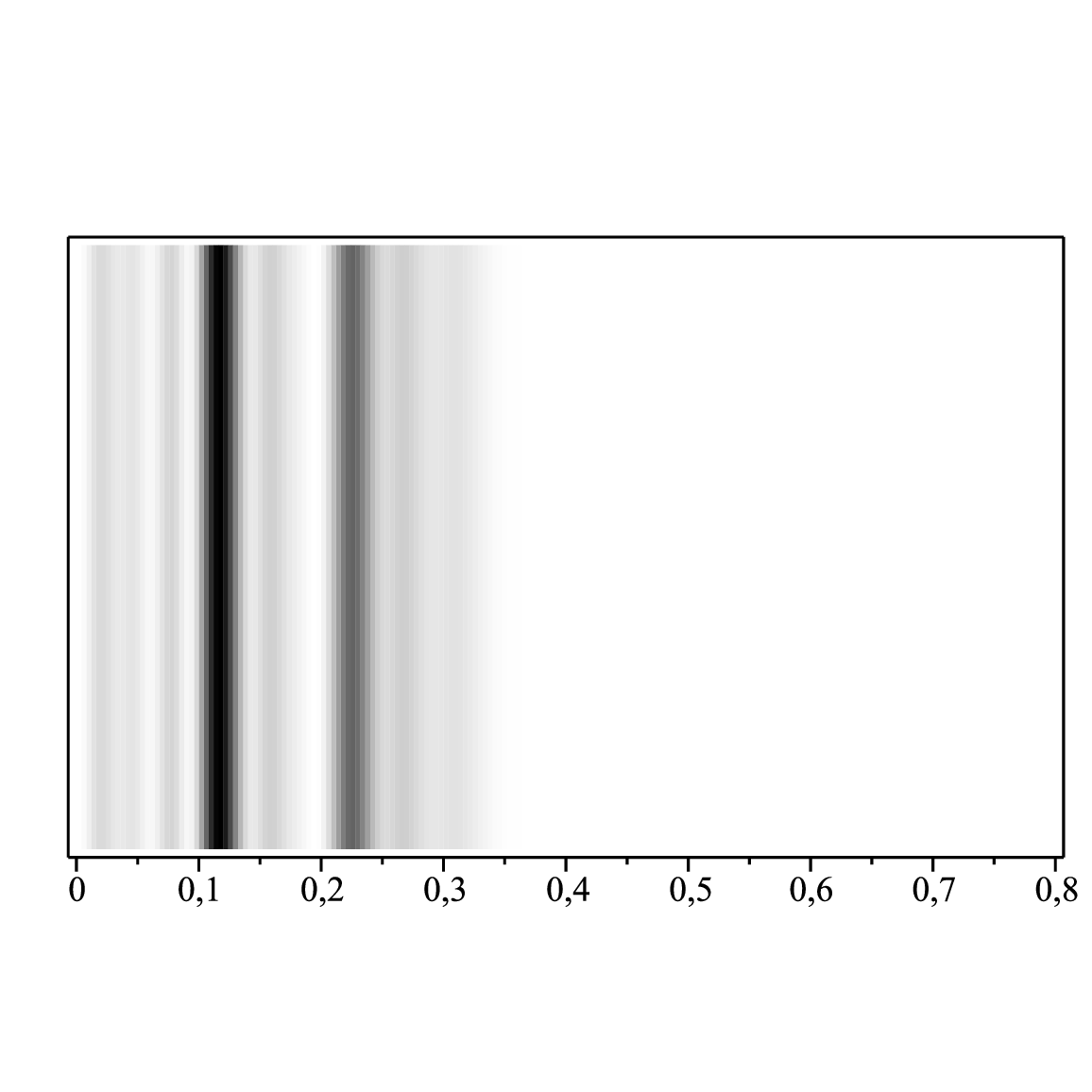}
\centerline{(b)}
\end{minipage}
\end{center}
\caption{The probability densities in (a) $t = 0$ and (b) $t = 1$, for $k = 1$ and $\sigma=-200$, respectively, obtained by the superposition of the first 10 eigenstates, where we consider $C_n = 1$ for $1 \leq n \leq 10$ and $C_n = 0$ to $n > 10$. Here the darker regions have the highest probability densities.}
\label{densidade}
\end{figure}

Using the wave functions, solution to the Schr\"{o}dinger Eq. (\ref{sec2eq14}), it is possible to obtain the expected value for the scale factor $<a>$ of this model through,

\begin{equation}
<a>(\tau)=\frac{\int_{0}^{\infty}a |\Psi(a,\tau)|^2da}{\int_{0}^{\infty}|\Psi(a,\tau)|^2da}\quad.
\label{scalefactor}
\end{equation}

We have studied here some cases where we vary $\sigma$ and ${\cal N}$. The result shows a universe oscillating between minimum and maximum values always greater than zero. The oscillation is due to the fact that we have bound states to the potencial and $<a>$. On the other hand, the minimum value of $<a>$ never goes to zero because of the repulsive force due the quantum effects for $a\rightarrow 0$.  Thus, we see that quantum effects eliminate existing singularities in the model, since the scale factor average value is never zero. For a fixed value of ${M}$ it is possible to observe that the larger the value of $\sigma$, the lower the average value of the scale factor and its amplitude. We also observed that the lower the value of the parameter $\sigma$ produces a higher oscillation frequency of the average value of the scale factor of the universe. We can see an example of these results in the Figure \ref{fxmedio}. 

If we set the value of $\sigma$ and increase the ${M}$, we observe that the mean value of the scale factor increases. In fact, the highter  the number of eigenstates considered in package construction, the greather the average energy of this package and consequently the greather amplitude for the expected value $<a>$. This result is shown in the Figure \ref{fxmedio2}.

\begin{figure}[h!]
\begin{center}
\begin{minipage}{0.45\linewidth}
\includegraphics[height=5.5cm,width=5.5cm]{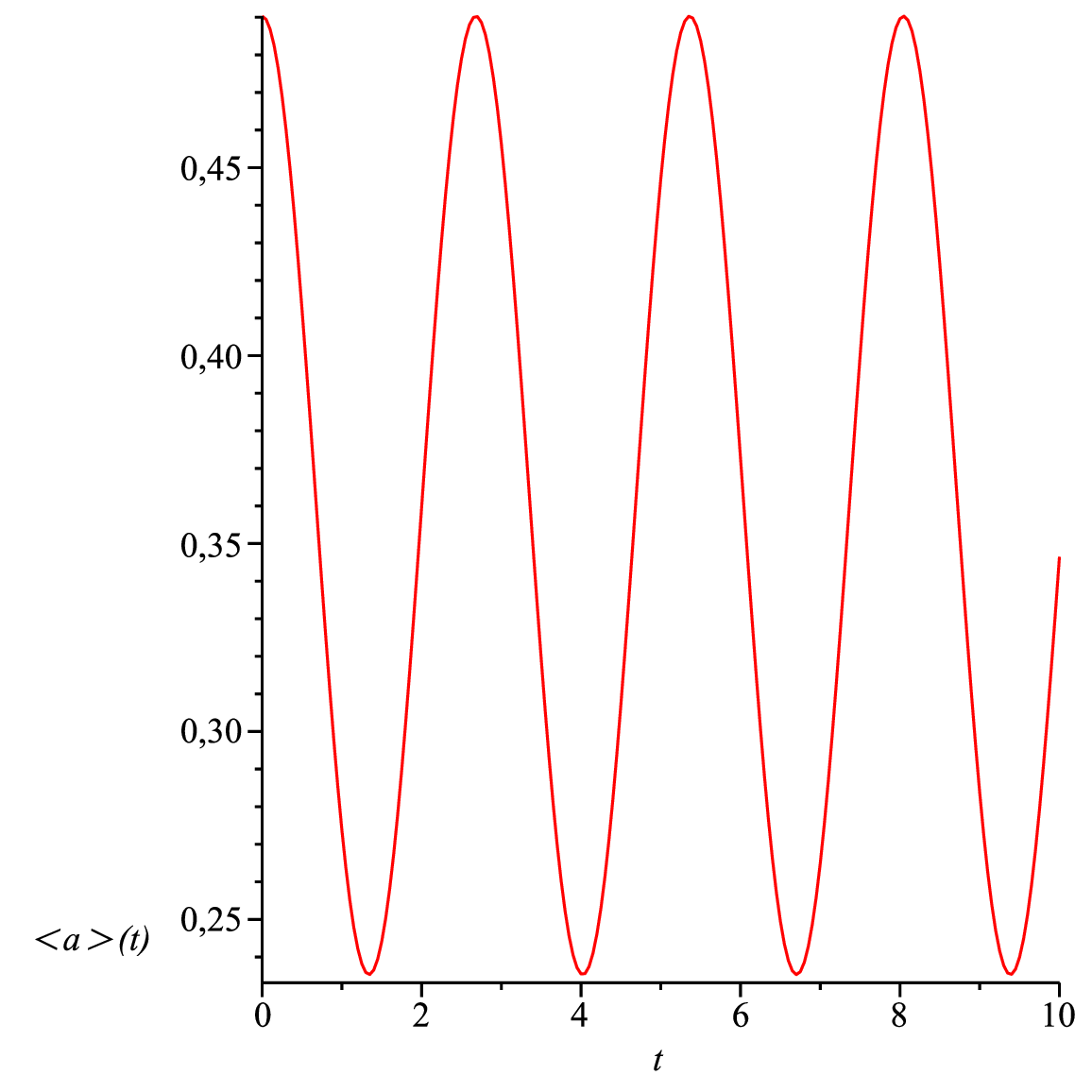}
\centerline{(a)}
\end{minipage}
\begin{minipage}{0.45\linewidth}
\includegraphics[height=5.5cm,width=5.5cm]{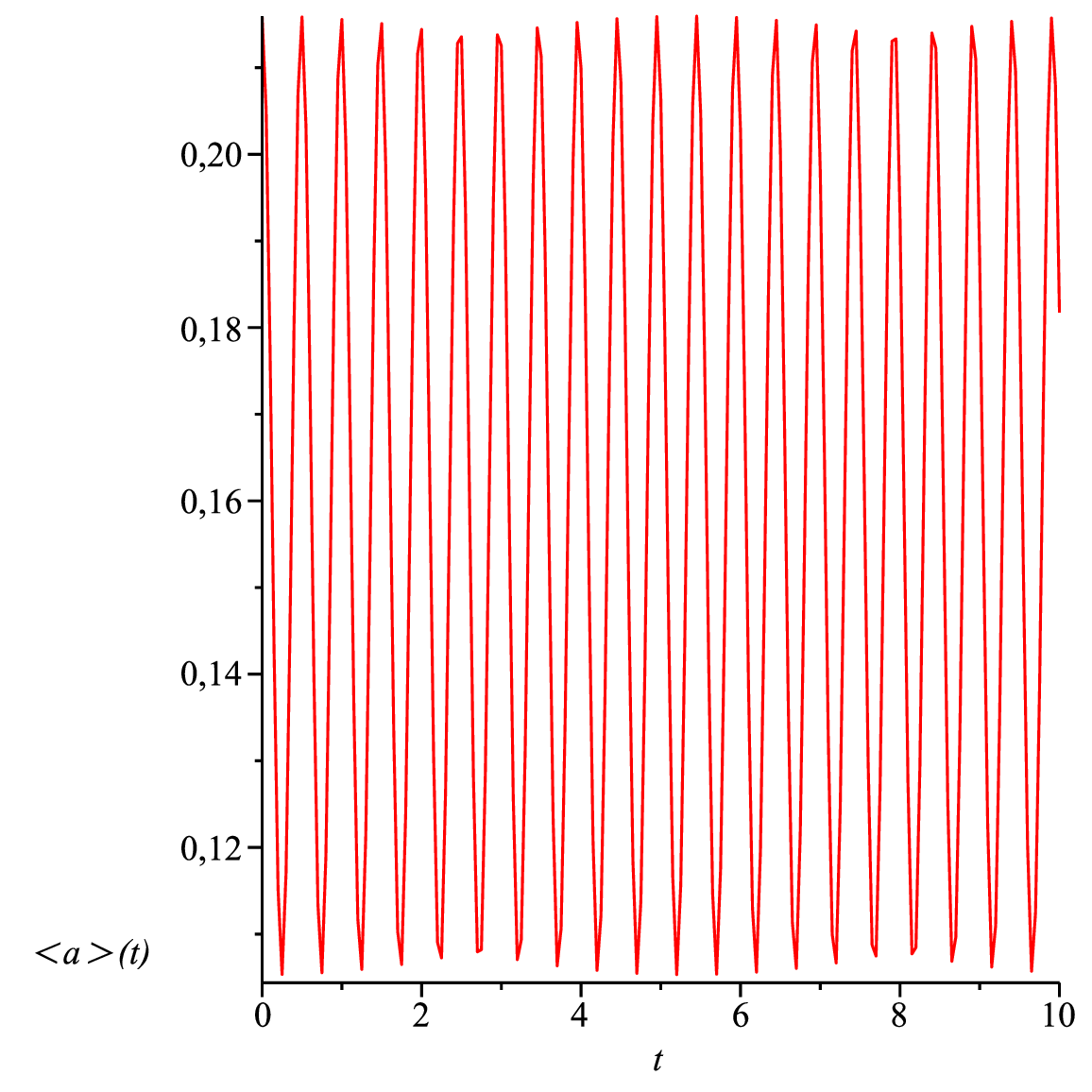}
\centerline{(b)}
\end{minipage}
\end{center}
\caption{Behaviour of the average value of the scale factor $\left<a\right>$ as a function of time. Here the wave packets used are obtained with $C_n = 1$ for $1\leq n \leq 10$ and $C_n = 0$ to $n > 10$.  In (a) temos $\sigma=-1$ and (b) $\sigma=-50$.}
\label{fxmedio}
\end{figure}

\begin{figure}[h!]
\begin{center}
\begin{minipage}{0.45\linewidth}
\includegraphics[height=5.5cm,width=5.5cm]{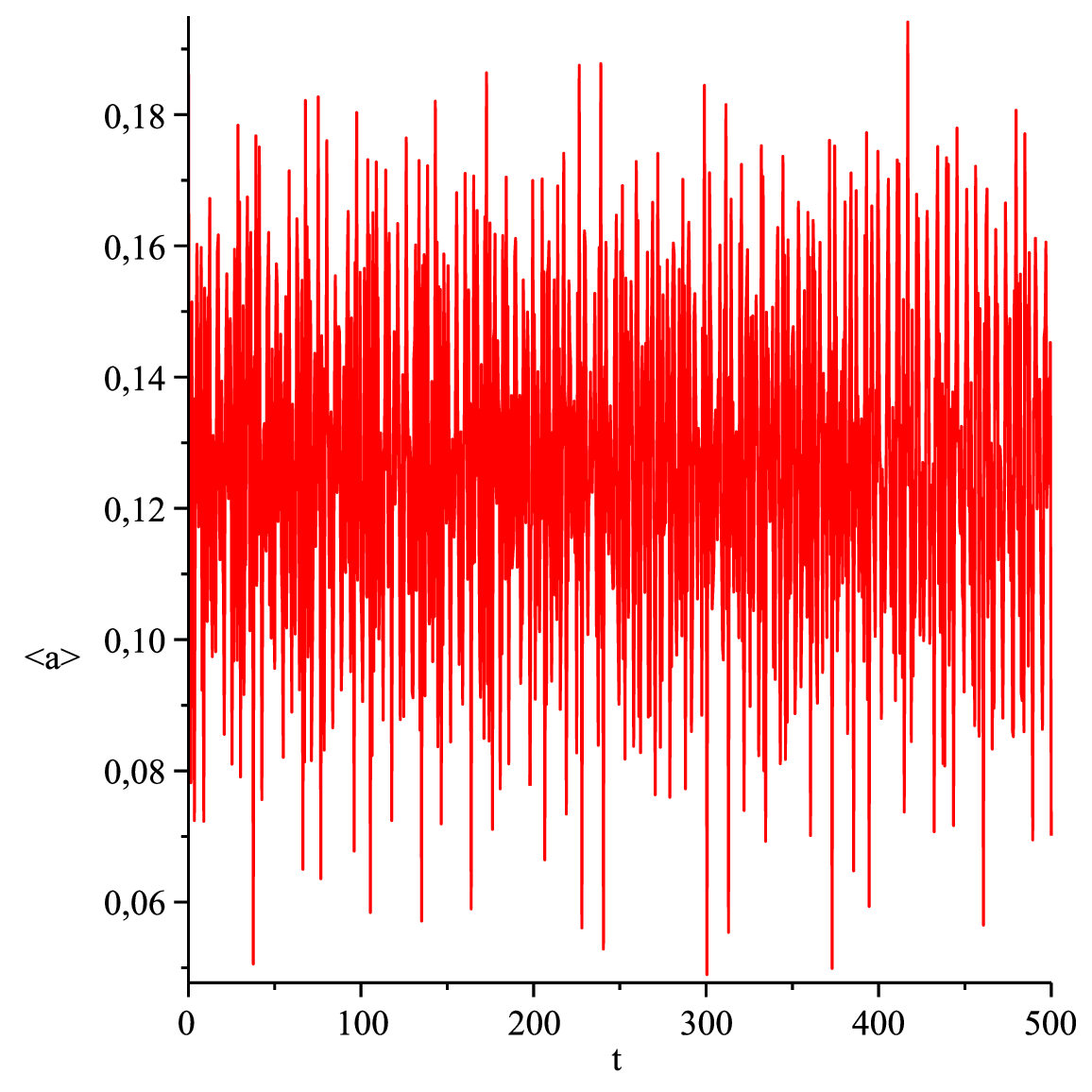}
\centerline{(a)}
\end{minipage}
\begin{minipage}{0.45\linewidth}
\includegraphics[height=5.5cm,width=5.5cm]{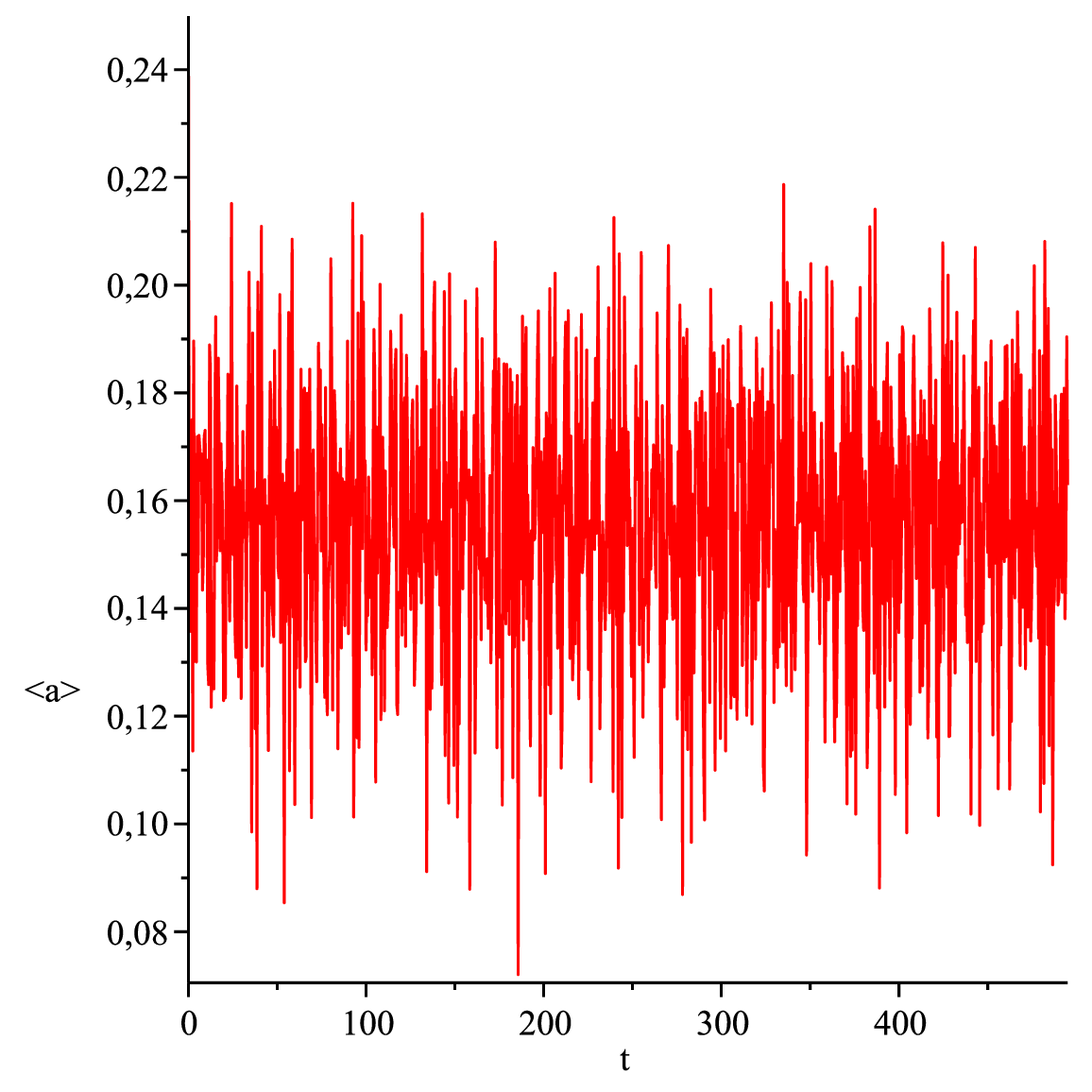}
\centerline{(b)}
\end{minipage}
\end{center}
\caption{Behaviour of the average value of the scale factor $\left<a\right>$ as a function of time. Here the wave packets used are obtained with: (a) $C_n = 1$ for $1\leq n \leq 5$ and $C_n = 0$ to $n > 5$ and (b) $C_n = 1$ for $1\leq n \leq 10$ and $C_n = 0$ to $n > 10$. Here we consider in both cases $\sigma = -200$.}
\label{fxmedio2}
\end{figure}
\begin{figure}[h!]
\begin{center}
\begin{minipage}{0.45\linewidth}
\includegraphics[height=5cm,width=5.2cm]{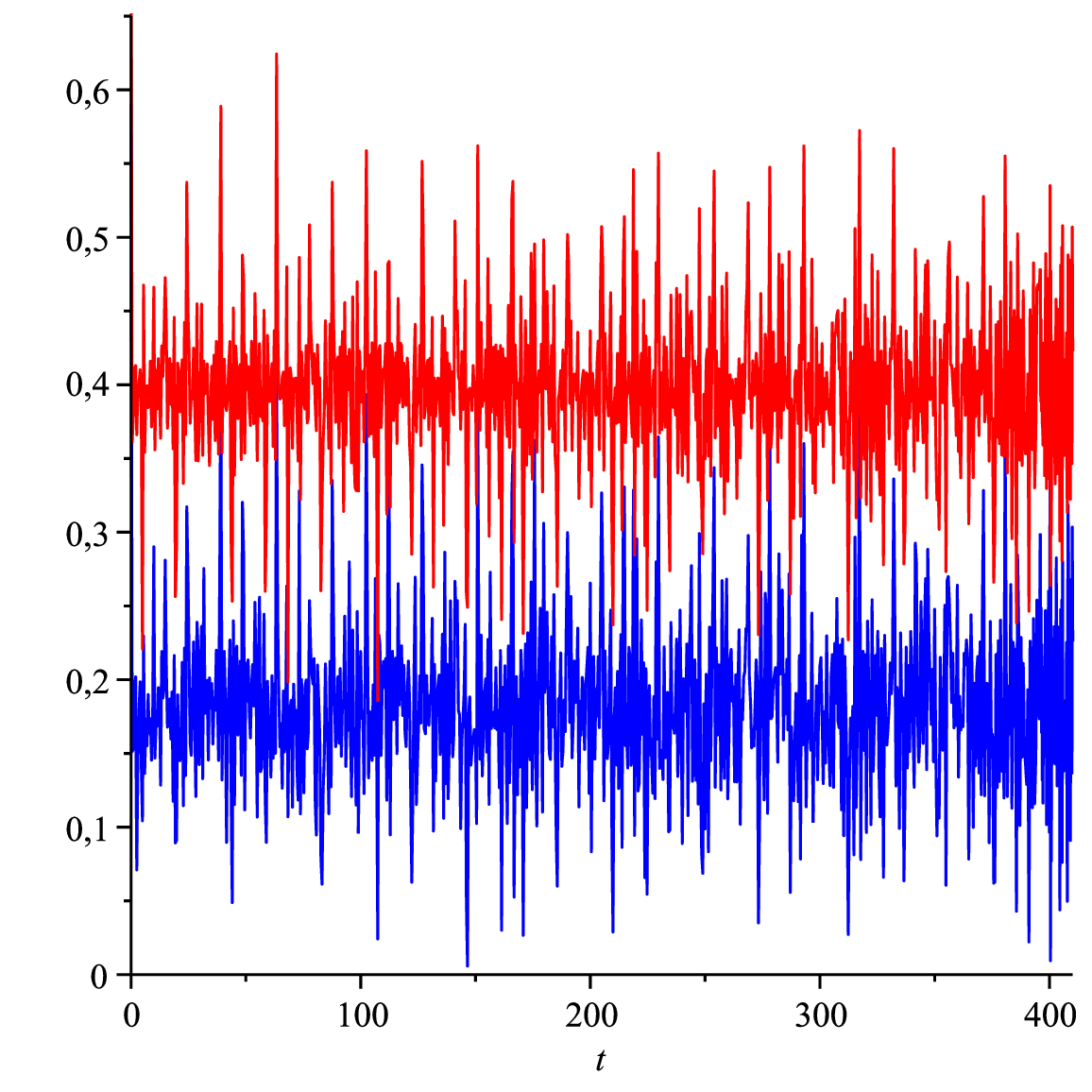}
\centerline{(a)}
\end{minipage}
\begin{minipage}{0.45\linewidth}
\includegraphics[height=5cm,width=5.2cm]{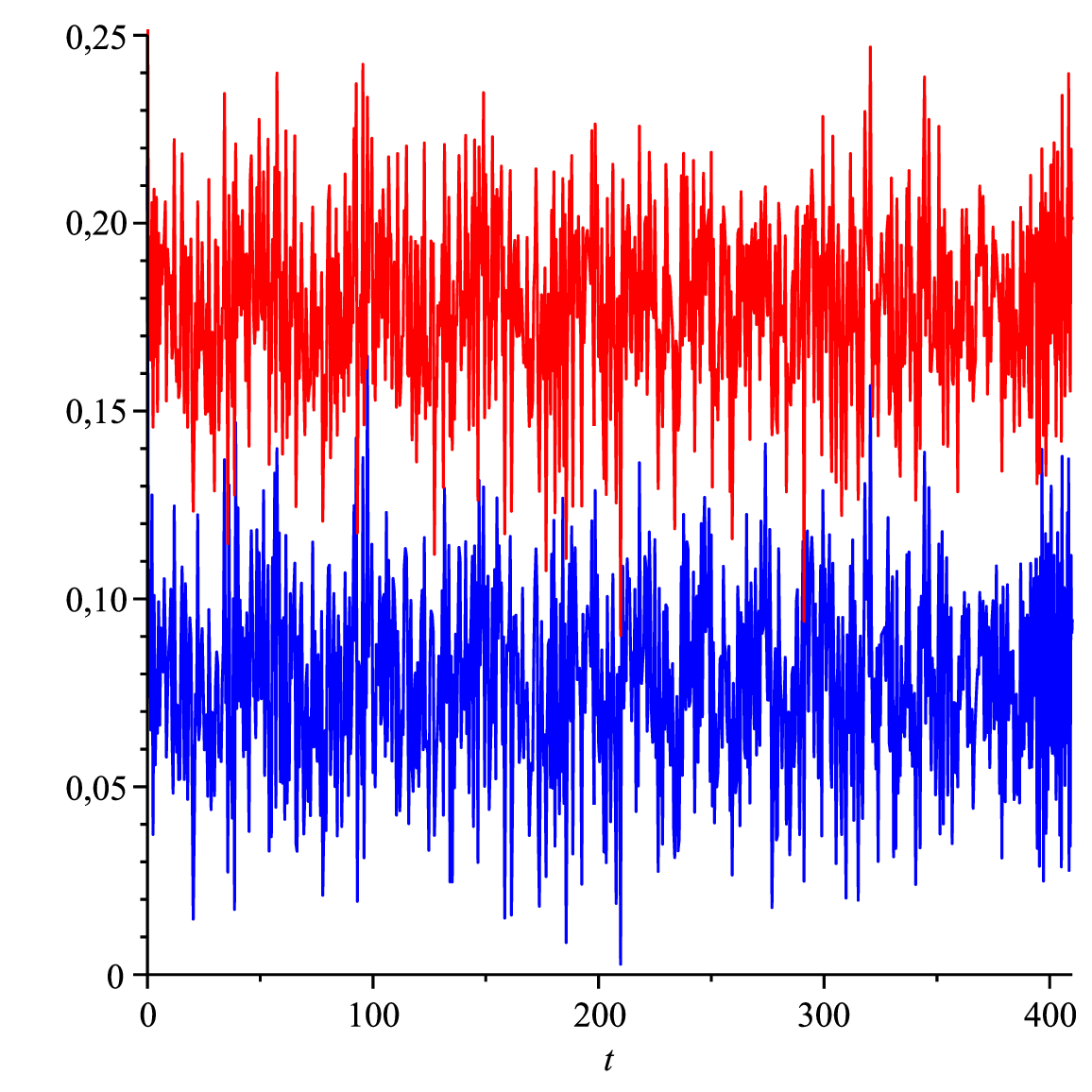}
\centerline{(b)}
\end{minipage}
\end{center}
\caption{Behaviour of the average value of the scale factor $\left<a\right>$ as a temporal function (in red) and the uncertainties (in blue) $\left<a\right>(\tau)- \Sigma(\tau)$. In both cases the wave packets used for these calculations were obtained by the superposition of the 15 lower levels. The system parameters here assume the values: (a) $\sigma=-6$, (b) $\sigma=-200$.}
\label{fincerteza2}
\end{figure}

The uncertainties $\Sigma(\tau)$ associated with the expected value of the scale factor of the universe ($\left<a\right>$), 
are defined by,
\begin{equation}
\label{20}
\Sigma(\tau) = \sqrt{\left<a^2\right> - \left<a\right>^2},
\end{equation}
where,
\begin{equation}
\label{21}
\left<a^2\right> = \frac{\int_{0}^{\infty}a^2\,|\Psi (a,\tau)|^2 da} {
\int_{0}^{\infty}|\Psi (a,\tau)|^2 da},
\end{equation}
and $\left<a\right>^2$ is given by the square of Eq. (\ref{scalefactor}).

We calculate those uncertainties for many different values of $\sigma$, ${\cal N}$ and time values. They are always positive. This means that even subtracting a standard deviations, the average value of the scale factor of the universe does not go to zero. This result is a strong indication that at the quantum level such models are free of singularities.

\section{Conclusions}
\label{CC}
In the present paper we studied the dynamics of a primordial universe filled with BEC and a radiation perfect fluid, using quantum cosmology. The universe has a Friedman-Lemaitre-Robertson-Walker geometry and the spatial sections have constant positive curvatures. In particular, we wanted to determine if the quantum description removes the singularities present in the classical model. 

In our model, the polytropic component for the pressure $p=\sigma \rho^2$ has a determining role in the evolution of the universe. For  polytropic constant $\sigma < 0$ that represents an attractive self-interaction the universe is bounded.

Classically, there are different possibilites for the evolution in time of the universe. In the range $-1020 \leq\sigma \leq -27$ the universe exhibits bouncing behavior or Big Crunch depending on the initial conditions imposed.  Outside this range the cosmological solutions are always Big Crunch type.

At the quantum level we solve the Wheeler-DeWitt equation through Galerkin's spectral method. The time variable was introduced phenomenologically using the degrees of freedom of radiation fluid which allows to obtain a
Hamiltonian constraint linear in one of the momenta. So, the Wheeler-DeWitt
equation can be reduced to a Schr\"odinger like equation.  We compute energy spectrum, eigenfunctions, wave packets and expected values for the scalar factor of the universe. The scale factor expected value oscillates between maximum and minimum values and never goes to zero. Therefore, this cosmological model is free from singularities, at the quantum level. We improved this result by showing that the quantity $\left<a\right>-\Sigma_a$ is always positive for many different wave packets, where $\Sigma_a$ stands for the standard deviation of $a$.

It would be worthwhile to extend these above results to other cases of cosmological interest. We can analyze, for example, the gravitational repulsive self-attractive case, where the $\sigma$ constant is positive ($\sigma > 0$). In this case, the effective potential $V_{eff}(a)$ diverge for $a = 1$. Thus, the situation is more complex and subtle that the gravitational attractive self-interaction case. We postpone a detailed study of such important problem for a future work.

\section*{Acknowledgment}

We thank FAPES for partial financial support. 


\end{document}